\RequirePackage{ifpdf}
\ifpdf 
\documentclass[pdftex]{sigma}
\else
\documentclass{sigma}
\fi

\begin{document}
\allowdisplaybreaks

\renewcommand{\PaperNumber}{005}

\FirstPageHeading

\ShortArticleName{Integrable Discrete Equations}

\ArticleName{Integrable Discrete Equations Derived by Similarity
Reduction of the Extended Discrete KP Hierarchy}

\Author{Andrei K. SVININ} 
\AuthorNameForHeading{A.K. Svinin}

\Address{Institute for System Dynamics and Control Theory,\\
 134 Lermontova Str., P.O. Box 1233, Irkutsk, 664033 Russia}

\Email{\href{mailto:svinin@icc.ru}{svinin@icc.ru}}

\ArticleDates{Received November 16, 2005, in final form January
08, 2006; Published online January 19, 2006}

\Abstract{We consider the extended discrete KP hierarchy and show
that similarity reduction of its subhierarchies lead to purely
discrete equations with dependence on some number of parameters
together with equations governing deformations with respect to
these para\-meters. It is written down discrete equations which
naturally generalize the first discrete Painlev\'e equation
$\mathrm{dP}_{\rm I}$ in a sense that autonomous version of these
equations admit the limit to the first Painlev\'e equation. It is
shown that each of these equations describes B\"acklund
transformations of Veselov--Shabat periodic dressing lattices with
odd period known also as Noumi--Yamada systems of type
$A_{2(n-1)}^{(1)}$.}


\Classification{37K10}

\section{Introduction}

The main goal of the present paper is to exhibit an approach,
which aims to construct a broad community of one-dimensional
discrete systems over finite number of fields sharing the property
of having Lax pair. It is achieved by similarity reduction of
equations of the so-called extended discrete KP (edKP) hierarchy
\cite{sv2,sv1,sv5}, which itself is proved to be a basis for
construction of many classes of integrable differential-difference
systems (see, for example, \cite{sv5} and references therein). By
analogy with the situation when Painlev\'e equations and its
hierarchies arise as a~result of similarity reduction of
integrable partial differential evolution equations
\cite{airault,flaschka}, discrete equations (like ${\rm dP}_{\rm
I}$) and its corresponding hierarchies also can be obtained by
application of group-theoretic methods.

In this connection it is necessary to mention various approaches
aiming to select integrable purely discrete one-dimensional
equations, for instance, by applying the confinement singularity
test~\cite{gr1}, from B\"acklund transformations for the
continuous Painlev\'e equations \cite{fokas1,nijhoff}, by imposing
integrable boundary conditions for two-dimensional discrete
equations and reductions to one-dimensional ones 
\cite{kazakova,gram,gram1,nijhoff1}, deriving discrete systems from proper
chosen representations of some affine Weyl group~\cite{noumi1}.

In the article we propose a scheme for constructing purely
discrete equations by similarity reductions of equations of edKP
hierarchy and then investigate more specific cases. Namely we
analyze one-component generalizations of ${\rm dP}_{\rm I}$ ---
discrete equations which turn out to serve as discrete symmetry
transformation for Veselov--Shabat periodic dressing lattices with
odd period. The latter also known in the literature as
Noumi--Yamada systems of type $A_{l}^{(1)}$ due to series of works
by Noumi with collaborators (see, for example,~\cite{noumi2})
where they selected the systems of ordinary differential equations
admitting a number of discrete symmetries which realized as
automorphisms on field of rational functions of corresponding
variables and constitute some representation of extended affine
Weyl group $\tilde{W}(A_{l}^{(1)})$.

The paper is organized as follows. In Section~2, we give necessary
information on Darboux-KP chain, its invariant manifolds and the
edKP hierarchy. In Section~3, we show that self-similar ansatzes
yield a large class of purely discrete systems supplemented by
equations governing deformations with respect to parameters
entering these systems. In Section~4, we show one-component
discrete equations naturally generalizing $\mathrm{dP}_{\rm I}$.
We prove that these discrete equations together with deformation
equations are equivalent to Veselov--Shabat periodic dressing
chains. Finally, in this section we show that each of these
discrete systems or more exactly its autonomous version has
continuous limit to $\mathrm{P}_{\rm I}$.

\section{Darboux-KP chain and edKP hierarchy}

\subsection{Darboux-KP chain and its invariant submanifolds}

In this section we give a sketch of the edKP hierarchy on the
basis of approach using the notion of Darboux-KP (DKP) chain
introduced in \cite{mpz}. Equations of DKP chain are defined in
terms of two bi-infinite sets of formal Laurent series. The first
set
\[
\left\{h(i)=z+\sum_{s=2}^{\infty}h_s(i)z^{-s+1} :
i\in\mathbb{Z}\right\}
\]
consists of generating functions for Hamiltonian densities of KP
hierarchy, and second one
\[
\left\{a(i)=z+\sum_{s=1}^{\infty}a_s(i)z^{-s+1} :
i\in\mathbb{Z}\right\}
\]
is formed by Laurent series $a(i)$ each of which relates two
nearest neighbors $h(i)$ and $h(i+1)$ by Darboux map
$h(i)\rightarrow h(i+1)=h(i)+a_x(i)/a(i)$. The DKP chain can be
interpreted as a~result of successive iterations of Darboux map
applying to any fixed solution of KP hierarchy, say $h(0)$, in
forward and backward directions. It is given by two equations
\begin{gather}
\partial_k h(i) = \partial H^{(k)}(i),  \nonumber\\
\partial_k a(i) = a(i)(H^{(k)}(i+1) - H^{(k)}(i)), \qquad
\partial_k\equiv \partial/\partial t_k, \qquad \partial\equiv \partial/\partial x
\label{DKP}
\end{gather}
first of which defines evolution equations of KP hierarchy in the
form of local conservation laws, and the second one serves as
compatibility condition of KP flows with Darboux map. The Laurent
series $H^{(k)}=H^{(k)}(i)$ in (\ref{DKP}) is the current of
corresponding conservation law constructed as special linear
combination over Fa\`a di Bruno polynomials
$h^{(k)}=(\partial+h)^k(1)$ by requiring to be projection of $z^k$
on ${\cal H}_{+}=\langle 1, h, h^{(2)},\ldots\rangle$~\cite{ca2}.

In \cite{sv1,sv5} we have exhibited two-parameter class of
invariant submanifolds of DKP chain $\mathcal{S}_l^n$ each of
which is specified by condition
\begin{gather}
z^{l-n+1}a^{[n]}(i)\in{\cal H}_{+}(i)\qquad \forall\,
i\in\mathbb{Z}. \label{byrelation}
\end{gather}
Here $a^{[k]}=a^{[k]}(i)$ is the discrete Fa\`a di Bruno iteration
calculated with the help of the recurrence relation $a^{[k+1]}(i)
= a(i)a^{[k]}(i+1)$ with $a^{[0]}\equiv 1$.

The invariant submanifolds $\mathcal{S}_l^1$ were presented
in~\cite{mpz}. It was shown there that the restriction of DKP
chain on $\mathcal{S}_0^1$ is equivalent to discrete KP hierarchy
\cite{ueno}. Restriction of DKP chain on $\mathcal{S}_0^1$,
$\mathcal{S}_0^2$, $\mathcal{S}_0^3$ and so on leads to the notion
of edKP hierarchy which is an infinite collection of discrete KP
like hierarchies attached to multi-times
$t^{(n)}\equiv(t^{(n)}_1,t^{(n)}_2,\ldots)$ with $n\geq 1$. All
these hierarchies ``live'' on the same phase-space ${\cal M}$
which can be associated with hyperplane whose points are
parametrized by infinite number of functions of discrete variable:
$\{a_k=a_k(i) : k\geq1\}$. One can treat these functions as
analytic ones whose the domain of definition is restricted to
$\mathbb{Z}$. We have shown in~\cite{sv1} that the flows on
$\mathcal{S}_0^n$ are given in the form of local conservation laws
as
\begin{gather}
z^{k(n-1)}\partial_k^{(n)}\ln
a(i)=K_{(n)}^{[kn]}(i+1)-K_{(n)}^{[kn]}(i), \label{::}
\end{gather}
that is on $\mathcal{S}_0^n$ one has
$H^{(k)}=z^{k(1-n)}K_{(n)}^{[kn]}$ with
\[
K_{(n)}^{[pn]}\equiv
\sum_{k=0}^pq_k^{(n,pn)}z^{k(n-1)}a^{[(p-k)n]}.
\]
Here the coefficients $q_k^{(n,r)}$ are some polynomials in
variables $\{a_l^{(s)}\equiv\Lambda^s(a_l)\}$ (for definition, see
below (\ref{:})). In what follows, we refer to (\ref{::}) as $n$th
subhierarchy of edKP hierarchy. In the following subsection we
show Lax pair for edKP hierarchy and write down evolution
equations on the functions $q_k^{(n,r)}$ generated by (\ref{::})
in its explicit form.

\subsection{Lax representation for edKP hierarchy}

First, let us recall the relationship between Fa\`a di Bruno
iterations and formal Baker--Akhiezer function of KP hierarchy
\cite{ca2,mpz} $\psi=\psi_i=\left(1+\sum\limits_{s\geq
1}w_s(i)z^{-s}\right)e^{\xi(t, z)}$ with $\xi(t,
z)=\sum\limits_{k\geq 1}t_kz^k$. One has
\[
h^{(k)}=\frac{\partial^k(\psi)}{\psi},\qquad
H^{(k)}=\frac{\partial_k(\psi)}{\psi},\qquad
a^{[k]}(i)=\frac{\Psi_{i+k}}{\Psi_i},\qquad \Psi_i\equiv
z^i\psi_i.
\]
These relations allow representing the DKP chain constrained by
relation (\ref{byrelation}) as compatibility condition of
auxiliary linear systems. As was shown in \cite{sv1,sv5}, the
restriction of DKP chain on~$\mathcal{S}_0^n$ leads to linear
discrete systems (see also~\cite{sv2})
\begin{gather}
Q_{(n)}^r\Psi = z^r\Psi,\qquad
z^{k(n-1)}\partial_k^{(n)}\Psi=(Q_{(n)}^{kn})_{+}\Psi \label{aux1}
\end{gather}
with the pair of discrete operators
\[
Q_{(n)}^r \equiv\sum_{s\geq
0}q_s^{(n,r)}z^{s(n-1)}\Lambda^{r-sn},\qquad (Q_{(n)}^{kn})_{+}
\equiv\sum_{s=0}^kq_s^{(n,kn)}z^{s(n-1)}\Lambda^{(k-s)n}.
\]
Here $\Lambda$ is a usual shift operator acting on arbitrary
function $f=f(i)$ of the discrete variable as $(\Lambda
f)(i)=f(i+1)$. The coefficients $q_s^{(n,r)}=q_s^{(n,r)}(i)$ are
uniquely defined as polynomial functions on ${\cal M}$ in
coordinates $a_l$ through the relation\footnote{As was mentioned
above, in fact, these are polynomials in $a_l^{(s)}$. For
notational convenience we assume that $q_0^{(n,r)}\equiv 1$ for
all possible $n$ and $r$.}
\begin{gather}
z^r=\sum_{s\geq 0}z^{s(n-1)}q_s^{(n,r)}a^{[r-sn]},\qquad
r\in\mathbb{Z}. \label{:}
\end{gather}
We assign to $a_l^{(s)}$ its scaling dimension: $[a_l^{(s)}]=l$.
One says that any polynomial $Q_k$ in $a_l$ is a homogeneous one
with degree $k$ if $Q_k\rightarrow \epsilon^k Q_k$ when
$a_l\rightarrow \epsilon^la_l$.

The consistency condition for the pair of equations (\ref{aux1})
reads as Lax equation
\[
z^{k(n-1)}\partial_k^{(n)}Q_{(n)}^r = [(Q_{(n)}^{kn})_{+},
Q_{(n)}^r]
\]
and can be rewritten in its explicit form as
\begin{gather}
\partial_k^{(n)}q_s^{(n, r)}(i)=Q_{s,k}^{(n,r)}(i)
\equiv\sum_{j=0}^{k}q_j^{(n, kn)}(i)\cdot q_{s-j+k}^{(n,r)}(i+(k-j)n)\nonumber\\
\phantom{\partial_k^{(n)}q_s^{(n,
r)}(i)=Q_{s,k}^{(n,r)}(i)\equiv{}}{} -\sum_{j=0}^{k}q_j^{(n,
kn)}(i+r-(s-j+k)n)\cdot q_{s-j+k}^{(n,r)}(i). \label{expl.form}
\end{gather}
It is important also to take into account algebraic relations
\begin{gather}
q_k^{(n,r_1+r_2)}(i)=\sum_{j=0}^{k}q_j^{(n,r_1)}(i)\cdot
q_{k-j}^{(n,r_2)}(i+r_1-jn) =\sum_{j=0}^{s}q_j^{(n,r_2)}(i)\cdot
q_{k-j}^{(n,r_1)}(i+r_2-jn) \label{r}
\end{gather}
coded in permutability operator relation $ Q_{(n)}^{r_1+r_2} =
Q_{(n)}^{r_1}Q_{(n)}^{r_2} = Q_{(n)}^{r_2}Q_{(n)}^{r_1}. $

It is quite obvious that equations (\ref{expl.form}) and (\ref{r})
admit reductions with the help of simple conditions
$q_s^{(n,r)}\equiv 0$ $(\forall\, s\geq l+1)$ for some fixed
$l\geq 1$. As was shown in \cite{sv1,sv5}, these reductions can be
properly described in geometric setting as double intersections of
invariant manifolds of DKP chain: $\mathcal{S}_{n,r,l} =
\mathcal{S}_0^n\cap\mathcal{S}_{l-1}^{ln-r}$. The formula
(\ref{expl.form}) when restricting to $\mathcal{S}_{n,r,l}$ is
proved to be a~container for many integrable
differential-difference systems (lattices) which can be found in
the literature (for reference see e.g.~\cite{sv5}).

\subsection{Conservation laws for edKP hierarchy}

The conserved densities for edKP hierarchy can be constructed in
standard way as residues
\begin{gather}
h_s^{(n)}={\rm Res}\,\big(Q_{(n)}^{sn}\big)=q^{(n,sn)}_s.
\label{Res}
\end{gather}
Corresponding currents are easily derived from (\ref{expl.form}).
One has
\begin{gather}
\partial_k^{(n)}q_s^{(n,sn)}(i)=Q_{s,k}^{(n,sn)}(i)=
J_{s,k}^{(n)}(i+n)-J_{s,k}^{(n)}(i)
=I_{s,k}^{(n)}(i+1)-I_{s,k}^{(n)}(i) \label{look}
\end{gather}
with
\[
J_{s,k}^{(n)}(i)=\sum_{l=0}^{k-1}\sum_{j=1}^{k-l}q_l^{(n,kn)}(i+(j+l-k-1)n)\cdot
q_{s+k-l}^{(n,sn)}(i+(j-1)n)
\]
and
\[
I_{s,k}^{(n)}(i)=\sum_{j=1}^n J_{s,k}^{(n)}(i+j-1)
=\sum_{l=0}^{k-1}\sum_{j=1}^{(k-l)n}q_l^{(n,kn)}(i+j+(l-k)n-1)\cdot
q_{s+k-l}^{(n,sn)}(i+j-1).
\]

We observe that by virtue of relations (\ref{r}) with $r_1=ln$,
$r_2=sn$ and $k=l+s$, the relation
$Q_{l,s}^{(n,ln)}=Q_{s,l}^{(n,sn)}$ with arbitrary $l,
s\in\mathbb{N}$ is identity and therefore we can write down
exactness property relation
\begin{gather}
\partial_s^{(n)}h_l^{(n)}=\partial_l^{(n)}h_s^{(n)}\qquad \forall\, s, l\geq 1.
\label{from}
\end{gather}

Looking at (\ref{look}), it is natural to suppose that there exist
some homogeneous polynomials $\xi_s^{(n)}$ in $q_k$ such that
\begin{gather}
h_s^{(n)}(i)=\sum_{j=1}^n\xi_s^{(n)}(i+j-1)\qquad \forall \, s\geq
1. \label{subst}
\end{gather}
If so, then one must recognize that $\{\xi_s^{(n)} : s\geq 1\}$ is
an infinite collection of conserved densities of $n$th
subhierarchy of edKP hierarchy, i.e.
\[
\partial^{(n)}_k\xi_s^{(n)}(i)=J_{s,k}^{(n)}(i+1)-J_{s,k}^{(n)}(i).
\]
Then the formula (\ref{subst}) says that each $\xi_s^{(n)}$ is
equivalent to $h_s^{(n)}/n$ modulo adding trivial densities.
Substituting (\ref{subst}) into (\ref{r}) with $r_1=1$ and
$r_2=kn$ leaves us with the following equation\footnote{Here and
in what follows $q_k(i)\equiv q_k^{(n,1)}(i)$. It is important to
note that the mapping $\{a_k\}\rightarrow\{q_k\}$ is invertible
and one can use $\{q_k\}$ as coordinates of ${\cal M}$.}:
\[
\xi_s^{(n)}(i+n)-\xi_s^{(n)}(i)= \sum_{j=1}^{s}q_j(i+jn)\cdot
q_{s-j}^{(n,sn)}(i)- \sum_{j=1}^{s}q_j(i)\cdot
q_{s-j}^{(n,sn)}(i+1-jn).
\]
One can check that solution of this equation is given by
\begin{gather}
\xi_s^{(n)}(i)= \sum_{l=1}^s\sum_{j=1}^lq_l(i+(j-1)n)\cdot
q_{s-l}^{(n,sn-1)}(i+1+(j-l-1)n). \label{xis}
\end{gather}
For example, we can write down the following:
\begin{gather*}
\xi_1^{(n)}(i) = q_1(i),\qquad
\xi_2^{(n)}(i)=q_2(i)+q_2(i+n)+q_1(i)\cdot\sum_{j=1-n}^{n-1}q_1(i+j),
\\
\xi_3^{(n)}(i)=q_3(i)+q_3(i+n)+q_3(i+2n)+
q_2(i)\cdot \sum_{j=1-2n}^{n-1}q_1(i+j)\\
\phantom{\xi_3^{(n)}(i)=}{} +q_2(i+n)\cdot
\sum_{j=1-n}^{2n-1}q_1(i+j)
+q_1(i)\cdot\left(\sum_{j=1-n}^{2n-1}q_2(i+j)\right.\\
\left.\phantom{\xi_3^{(n)}(i)=}{}
+\sum_{l=1-n}^{n-1}\sum_{j=1-n}^{n-1}q_1(i+l)\cdot q_1(i+j+l)
+\sum_{l=1}^{n-1}\sum_{j=l}^{n-1}q_1(i+l-n)\cdot q_1(i+j) \right).
\end{gather*}

Now let us show that conserved densities $\xi_s^{(n)}$ satisfy
some characteristic equations. From (\ref{Res}) one derives
\[
\frac{\delta h_s^{(n)}}{\delta q_l}(i)=
n\frac{\delta\xi_s^{(n)}}{\delta q_l}(i)=
snq_{s-l}^{(n,sn-1)}(i+1-ln).
\]
Substituting the latter in (\ref{xis}) one gets
\[
s\xi_s^{(n)}(i)=\sum_{l=1}^s\sum_{j=1}^lq_l(i+(j-1)n)
\frac{\delta\xi_s^{(n)}}{\delta q_l}(i+(j-1)n).
\]

\section[Self-similar ansatzes for edKP hierarchy
and integrable discrete equations]{Self-similar ansatzes for edKP hierarchy\\
and integrable discrete equations}

In what follows, we consider $n$th edKP subhierarchy restricted to
the first $p$ time variables $\{t_1^{(n)},\ldots, t_p^{(n)}\}$
with $p\geq 2$. This system is obviously invariant under the group
of scaling transformations
\[
G_p=\left\{g : q_k\rightarrow\epsilon^k q_k\;\; (k\geq 1),\;\;
t_l^{(n)}\rightarrow\epsilon^{-l}t_l^{(n)}\;\; (l=1,\ldots,
p)\right\}.
\]
Since $q_k^{(n,r)}$, $Q_{k,s}^{(n,r)}$, $\xi_k^{(n)}$ are
homogeneous polynomials in $q_k$ of corresponding degrees, the
transformation $g$ yields
\[
q_k^{(n,r)}\rightarrow\epsilon^kq_k^{(n,r)},\qquad
Q_{k,s}^{(n,r)}\rightarrow\epsilon^{k+s}Q_{k,s}^{(n,r)},\qquad
\xi_k^{(n)}\rightarrow\epsilon^k\xi_k^{(n)}.
\]

Let us consider similarity reductions of $n$th edKP subhierarchy
requiring
\begin{gather}
\left(k+\sum_{j=1}^pjt_j^{(n)}\partial_j^{(n)}\right)q_k=
kq_k+\sum_{j=1}^pjt_j^{(n)}Q_{k,j}^{(n,1)}=0\qquad \forall\, k\geq
1. \label{self-similarity}
\end{gather}
Then any homogeneous polynomial in $q_l$ satisfies corresponding
self-similarity condition. In particular, one has
\begin{gather}
\left(k+\sum_{j=1}^pjt_j^{(n)}\partial_j^{(n)}\right)q_k^{(n,r)}=
kq_k^{(n,r)}+\sum_{j=1}^pjt_j^{(n)}Q_{k,j}^{(n,r)}=0. \label{:::}
\end{gather}
More explicitly, one can rewrite (\ref{:::}) as
\begin{gather*}
q_k^{(n,r)}(i)\left(k+\alpha_i^{(n,n)}-\alpha_{i+r-kn}^{(n,n)}\right)
+\sum_{s=1}^pq_{k+s}^{(n,r)}(i+sn)\sum_{j=s}^pjt_j^{(n)}
q_{j-s}^{(n,jn)}(i)
\\
\qquad {}-\sum_{s=1}^pq_{k+s}^{(n,r)}(i)
\sum_{j=s}^pjt_j^{(n)}q_{j-s}^{(n,jn)}(i+r-(k+s)n) =0
\end{gather*}
with
\[
\alpha_i^{(n,n)}\equiv \sum_{j=1}^pjt_j^{(n)}h_j^{(n)}(i)=
\sum_{s=1}^n\alpha_{i+s-1}^{(n)},
\]
where $\alpha^{(n)}\equiv
\sum\limits_{j=1}^pjt_j^{(n)}\xi_j^{(n)}$. We observe that
$\alpha^{(n,n)}$ do not depend on evolution parameters
$\{t_1^{(n)},\ldots,t_p^{(n)}\}$ provided that
(\ref{self-similarity}) is valid. Indeed, taking into account
exactness property (\ref{from}) one has
\[
\partial_s^{(n)}\alpha^{(n,n)}=
sh_s^{(n)}+\sum_{j=1}^pjt_j^{(n)}\partial_s^{(n)}h_j^{(n)}=
\left(s+\sum_{j=1}^pjt_j^{(n)}\partial_j^{(n)}\right)h_s^{(n)}=0.
\]

For Baker--Akhiezer function, one has the corresponding
self-similarity condition in the form
\[
\left(z\partial_z-\sum_{j=1}^pjt_j^{(n)}\partial_j^{(n)}\right)\psi=0
\]
or
\[
z\partial_z\psi_i= \alpha^{(n,n)}_i\psi_i+
\sum_{s=1}^p\psi_{i+sn}\sum_{j=s}^pjt_j^{(n)}q_{j-s}^{(n,jn)}.
\]

One can rewrite the above formulas in terms of self-similarity
ansatzes
\begin{gather*}
T_l = \frac{t_l^{(n)}}{(pt_p^{(n)})^{l/p}},\qquad l = 1,\ldots,
p-1,\qquad \xi = (pt_p^{(n)})^{1/p}z,
\\
q_s^{(n,r)}=\frac{1}{(pt_p^{(n)})^{s/p}}x_s^{(n,r)},\qquad
Q_{s,k}^{(n,r)}=\frac{1}{(pt_p^{(n)})^{(s+k)/p}}X_{s,k}^{(n,r)},\qquad
\xi_s^{(n)}=\frac{1}{(pt_p^{(n)})^{s/p}}\zeta_s^{(n)}.
\end{gather*}

At this point we can conclude that similarity reduction of $n$th
edKP subhierarchy yields the system of purely discrete equations
\begin{gather}
x_k^{(n,r)}(i)\left(k+\alpha_i^{(n,n)}-\alpha_{i+r-kn}^{(n,n)}\right)
+\sum_{s=1}^{p-1}x_{k+s}^{(n,r)}(i+sn)
\left(\sum_{j=s}^{p-1}jT_jx_{j-s}^{(n,jn)}(i)+x_{p-s}^{(n,pn)}(i)\right)
\nonumber\\
\qquad{}-\sum_{s=1}^{p-1}x_{k+s}^{(n,r)}(i)\left(
\sum_{j=s}^{p-1}jT_jx_{j-s}^{(n,jn)}(i+r-(k+s)n)+x_{p-s}^{(n,pn)}(i+r-(k+s)n)
\right)
\nonumber\\
\qquad{}+x_{k+p}^{(n,r)}(i+pn)-x_{k+p}^{(n,r)}(i)=0, \label{odin}
\\
\alpha^{(n)}=\sum_{j=1}^{p-1}jT_j\zeta_j^{(n)}+\zeta_p^{(n)}
\label{dva}
\end{gather}
supplemented by deformation equations
\[
\partial_{T_k}x_s^{(n,r)}=X_{s,k}^{(n,r)}, \qquad k=1,\ldots, p-1.
\]
These equations appear as a consistency condition of the following
linear isomonodromy problem
\begin{gather*}
\xi\psi_{i+r}+
\sum_{k\geq1}\xi^{1-k}x_k^{(n,r)}\psi_{i+r-kn}=\xi\psi_i,
\\
\partial_{T_k}\psi_i=\xi^k\psi_{i+kn}+
\sum_{s=1}^k\xi^{k-s}x_s^{(n,kn)}\psi_{i+(k-s)n}, \qquad
k=1,\ldots, p-1,
\\
\xi\partial_{\xi}\psi_i=\alpha_i^{(n,n)}\psi_i+\sum_{s=1}^p\psi_{i+sn}
\left( \sum_{j=s}^{p-1}jT_jx_{j-s}^{(n,jn)}(i)+x_{p-s}^{(n,pn)}(i)
\right).
\end{gather*}

The pair of equations (\ref{odin}) and (\ref{dva}) when
restricting dynamics on $\mathcal{S}_{n,r,l}$ becomes a system
over finite number of fields. If one requires $x_k^{(n,
r)}(i)\equiv 0$ $(\forall\, k\geq l+1)$ then $l$-th equation in
(\ref{odin}) is specified as
\[
x_l^{(n, r)}(i)\cdot\left\{ l + \sum_{s=1}^n\alpha_{i+s-1}^{(n)} -
\sum_{s=1}^n\alpha_{i+r-ln+s-1}^{(n)} \right\} = 0.
\]
Since it is supposed that $x_l^{(n, r)}\not\equiv 0$ then the
constants $\alpha_i^{(n)}$ are forced to be subjects of constraint
\begin{gather}
l+\sum_{s=1}^n\alpha_{i+s-1}^{(n)} -
\sum_{s=1}^n\alpha_{i+r-ln+s-1}^{(n)} = 0. \label{tri}
\end{gather}
Remark that this equation makes no sense in the case $r=ln$. In
the following section we consider restriction on
$\mathcal{S}_{n,1,1}$ corresponding to Bogoyavlenskii lattice.

\section{Examples of one-field discrete equations}

\subsection[One-field discrete system generalizing $\mathrm{dP}_{\rm I}$ and its
hierarchies]{One-field discrete system generalizing
$\boldsymbol{\mathrm{dP}_{\rm I}}$ and its hierarchies}

Let us consider reductions corresponding to $\mathcal{S}_{n,1,1}$
with arbitrary $n\geq 2$. For $p=2$ one obtains one-field
system\footnote{Here $x_i\equiv x_1^{(n,1)}(i)$ and $T\equiv
T_1$.}
\begin{gather}
T+\sum_{s=1-n}^{n-1}x_{i+s}=\frac{\alpha_i^{(n)}}{x_i},\qquad
1+\sum_{s=1}^{n-1}\alpha^{(n)}_{i+s}-\sum_{s=1}^{n-1}\alpha^{(n)}_{i-s}
= 0 \label{dp3}
\end{gather}
together with deformation equation
\begin{gather}
x_i^{\prime} = x_i\left(\sum_{s=1}^{n-1}x_{i+s} -
\sum_{s=1}^{n-1}x_{i-s}\right) \label{bog}
\end{gather}
which is nothing else but Bogoyavlenskii lattice. The pair of
equations (\ref{dp3}) is a specification of (\ref{dva}) and
(\ref{tri}). In the case $n=2$, (\ref{dp3}) becomes
$\mathrm{dP}_{\rm I}$ while (\ref{bog}) turns into Volterra
lattice. Hierarchy of~$\mathrm{dP}_{\rm I}$ appears to be related
with a matrix model of two-dimensional gravity (see, for
example,~\cite{fokas} and references therein). In \cite{cr} Joshi and Cresswell 
found out representation of ${\rm dP}_{\rm I}$
hierarchy with the help of recursion
operator.

We are in position to exhibit hierarchy of more general equation
(\ref{dp3}) by considering the cases $p=3, 4$ and so on. From
(\ref{xis}) one has
\[
\zeta_s^{(n)}(i)=x_ix_{s-1}^{(n,sn-1)}(i+1-n).
\]
Substituting this into (\ref{dva}) leads to
\begin{gather}
T_1+\sum_{j=2}^{p-1}jT_jx_{j-1}^{(n,jn-1)}(i+1-n)+x_{p-1}^{(n,pn-1)}(i+1-n)=
\frac{\alpha_i^{(n)}}{x_i}, \label{chetyre}
\end{gather}
where $\alpha^{(n)}_i$'s are constants which are supposed to solve
the same algebraic equations as in (\ref{dp3}). As an example, for
$p=3$, one has
\[
T_1+2T_2\sum_{s=1-n}^{n-1}x_{i+s}+
\sum_{s=1-n}^{n-1}\sum_{j=1-n}^{n-1}x_{i+s}x_{i+j+s}
+\sum_{s=1}^{n-1}\sum_{j=s}^{n-1}x_{i+s-n}x_{i+j}=\frac{\alpha_i^{(n)}}{x_i}.
\]

Equation (\ref{chetyre}) should be complemented by deformation
ones
\[
\partial_{T_k}x_i=x_i\left(
\sum_{s=1}^{n-1}\zeta_k^{(n)}(i+s)-\sum_{s=1}^{n-1}\zeta_k^{(n)}(i-s)
\right),\qquad k=1,\ldots, p-1
\]
which are higher members in Bogoyavlenskii lattice hierarchy.

\subsection[Equivalence (\ref{dp3})
and (\ref{bog}) to Veselov-Shabat periodic dressing lattices]{Equivalence (\ref{dp3})
and (\ref{bog}) to Veselov--Shabat periodic dressing lattices}

Here we restrict ourselves by consideration only (\ref{dp3}) and
(\ref{bog}) without their hierarchies. Our aim is to show
equivalence of this pair of equations to well-known
Veselov--Shabat periodic dressing lattices. To this end, we
introduce the variables $\{r_0,\ldots, r_{2(n-1)}\}$ by
identifying
\begin{gather*}
x_{i+k-1}=-r_{2k}-r_{2k+1},\qquad
x_{i+n+k-2}=-r_{2k-1}-r_{2k},\qquad k=1,\ldots, n-1,\\
\sum_{s=1}^{2(n-1)}x_{i+s-1}+T=r_0+r_1,
\end{gather*}
where subscripts are supposed to be elements of
$\mathbb{Z}/(2n-1)\mathbb{Z}$. One can show that with the
constants
\begin{gather*}
c_{2k}=\alpha_{i+k-1}^{(n)},\qquad
c_{2k-1}=-\alpha_{i+k+n-2}^{(n)}, \qquad
k=1,\ldots, n-1,\\
c_0=1-\sum_{s=1}^{n-1}\alpha_{i+s-1}^{(n)}+\sum_{s=1}^{n-1}\alpha_{i+s+n-2}^{(n)}
\end{gather*}
the equations (\ref{dp3}) and (\ref{bog}) can be rewritten as
Veselov--Shabat periodic dressing chain \cite{veselov,adler}
\begin{gather}
r_k^{\prime}+r_{k+1}^{\prime}=r_{k+1}^{2}-r_k^{2}+c_k,\qquad
r_{k+N}=r_k,\qquad c_{k+N}=c_k \label{vsh}
\end{gather}
with $N=2n-1$. For $N=3$ this system is known to be equivalent to
${\rm P}_{\rm IV}$. In the variables $f_k=r_k+r_{k+1}$ the system
(\ref{vsh}) has the form \cite{veselov}
\begin{gather}
f_k^{\prime}=
f_k\left(\sum_{r=1}^{n-1}f_{k+2r-1}-\sum_{r=1}^{n-1}f_{k+2r}\right)+c_k,\qquad
k=0,\ldots, 2(n-1). \label{noumi}
\end{gather}

The system  (\ref{noumi}) for $n\geq 3$ can be considered as
higher-order generalization of ${\rm P}_{\rm IV}$ written in
symmetric form. It was investigated in \cite{noumi2} where Noumi
and Yamada showed its invariance with respect to discrete symmetry
transformations (or B\"acklund--Schlesinger transformations) which
are realized as a number of automorphisms $\{\pi, s_0,\ldots ,
s_{2(n-1)}\}$ on the field of rational functions in variables
$\{c_0,\ldots , c_{2(n-1)}\}$ and $\{f_0,\ldots , f_{2(n-1)}\}$ as
follows:
\begin{gather*}
s_k(c_k)=-c_k,\qquad s_k(c_l)=c_l+c_k, \qquad
l=k\pm 1,\qquad s_k(c_l)=c_l, \qquad l\neq k, k\pm 1,  \\
s_k(f_k)=f_k,\qquad s_k(f_l)=f_l\pm\frac{c_k}{f_k}, \qquad l=k\pm
1,\qquad
s_k(f_l)=f_l, \qquad l\neq k, k\pm 1, \\
\pi(f_k)=f_{k+1},\qquad \pi(c_k)=c_{k+1}.
\end{gather*}
One can easily rewrite the formulas of action of automorphisms on
generators $r_k$ as follows:
\begin{gather*}
s_k(r_k)=r_k-\frac{c_k}{r_k+r_{k+1}},\qquad
s_k(r_{k+1})=r_k+\frac{c_k}{r_k+r_{k+1}}, \\
s_k(r_l)=r_l,\qquad l\neq k, k+1,\qquad \pi(r_k)=r_{k+1}.
\end{gather*}

It is known by \cite{noumi1} that this set of automorphisms define
a representation of the extended affine Weyl group
$\tilde{W}(A_l^{(1)})=\langle \pi, s_0,\ldots , s_l\rangle$ with
$l=2(n-1)$ whose generators satisfy the relations
\[
s_k^2=1,\qquad s_ks_m=s_ms_k,\qquad m\neq k\pm 1,\qquad
(s_ks_m)^3=1,\qquad m=k\pm 1
\]
for $k, m=0,\ldots , l$ and
\[
\pi^{l+1}=1,\qquad \pi s_k=s_{k+1}\pi. \label{vtorym}
\]

The shift governed by discrete equations (\ref{dp3}) affects on
the system (\ref{vsh}) as B\"acklund transformation
\begin{gather*}
\overline{r}_0=r_2+\frac{c_1}{r_1+r_2},\qquad
\overline{r}_k=r_{k+2},\qquad k=1,\ldots , 2n-3, \\
\overline{r}_{2(n-1)}=r_1-\frac{c_1}{r_1+r_2}, \\
\overline{c}_0=c_1+c_2,\qquad \overline{c}_k=c_{k+2},\qquad k=1,\ldots , 2n-4, \\
\overline{c}_{2n-3}=c_0+c_1,\qquad
\overline{c}_{2(n-1)}=-c_{2(n-1)}.
\end{gather*}
By direct calculations one can check that it coincides with
$\tau=s_1\pi^2\in\tilde{W}\big(A_{2(n-1)}^{(1)}\big)$.

\subsection{Continuous limit of stationary version of (\ref{dp3})}

Let us show in this subsection that stationary version of the
equation~(\ref{dp3})
\begin{gather}
T+\sum_{s=1-n}^{n-1}x_{i+s}=\frac{\alpha}{x_i}, \label{dp4}
\end{gather}
for any $n$, is integrable discretization of $\mathrm{P}_{\rm I}$:
$w^{\prime\prime} = 6w^2 + t$.

One divides the real axis into segments of equal length
$\varepsilon$. One considers discrete set of values $\{t=i\varepsilon :
i\in\mathbb{Z}\}$. Values
of the function $w$, respectively, are taken for all such values
of the variable $t$. Therefore one can denote $w(t) = w_i$. Let
\begin{gather}
x_i = 1 + \varepsilon^2w_i,\qquad
\alpha=1-2n-\varepsilon^4t,\qquad T_1 = -2n+1. \label{s}
\end{gather}
Substituting (\ref{s}) in the equation (\ref{dp4}) and taking into
account the relations of the form
\[
x_{i+1} = 1 + \varepsilon^2w_{i+1} = 1 + \varepsilon^2\left\{w +
\varepsilon w^{\prime} + \frac{\varepsilon^2}{2}w^{\prime\prime} +
\cdots \right\}
\]
and turning then $\varepsilon$ to zero we obtain, in continuous
limit, the equation
\[
\sum_{s=1}^{n-1}(n-s)^2\cdot w^{\prime\prime} = - t - (2n-1)w^2
\]
which, by suitable rescaling, can be deduced to canonical form of
$\mathrm{P}_{\rm I}$. This situation goes in parallel with that
when Bogoyavlenskii lattice (\ref{bog}), for all values of $n\geq
2$, has continuous limit to Korteweg--de Vries equation~\cite{bog}.

\section{Conclusion}

We have considered in the paper edKP hierarchy restricted to $p$ evolution
parameters $\{t_1^{(n)},\ldots$, $t_p^{(n)}\}$. It was shown that self-similarity
constraint imposed on this system is equivalent to purely discrete 
equations~(\ref{odin}) and~(\ref{dva}) supplemented by $(p-1)$ deformation equations
which in fact are evolution equations governing $(p-1)$ flows of $n$th edKP
subhierarchy.
What is crucial in our approach is that we selected quantities $\alpha^{(n,n)}$
which enter these discrete equations and turn out to be independent on
evolution parameters due to exactness property  for conserved densities.

Discrete systems over finite number of fields arise when one restricts edKP
hierarchy on invariant submanifold ${\cal S}_{n,r,l}$. In the present paper
we considered only one-field discrete equations corresponding to
${\cal S}_{n,1,1}$ with $n\geq 2$. It is our observation that in this case
discrete equation under consideration supplemented by deformation one is
equivalent to Veselov--Shabat periodic dres\-sing lattice with odd period.
It is known due to Noumi and Yamada that this finite-dimensional system
of ordinary differential equations admits finitely generated group of
B\"acklund transformations which realizes representation of extended
affine Weyl group $\tilde{W}(A^{(1)}_{2(n-1)})$. It is natural to expect
that restriction of edKP hierarchy to other its invariant submanifolds
also can yield finite-dimensional systems invariant under some finitely
generated groups of discrete transformations. We are going to present
relevant results on this subject in subsequent publications.

\subsection*{Acknowledgements}

The author thanks the organizers of Sixth International Conference
``Symmetry in Nonlinear Mathematical Physics'' (June 20--26, 2005, Kyiv) for
hospitality and their work which allows to create friendly
atmosphere during this Conference. The present paper is the
written version of the talk delivered by the author at this
conference.

\LastPageEnding


\begin{thebibliography}{99}
\footnotesize

\bibitem{adler}
Adler V.E., Nonlinear chains and Painlev\'e equations, {\it
Phys.~D}, 1994, V.73, 335--351.

\bibitem{airault}
Airault H., Rational solutions of Painlev\'e equations, {\it Stud.
Appl. Math.}, 1979, V.61, 31--53.

\bibitem{bog}
Bogoyavlenskii O.I., Breaking solitons: nonlinear integrable
equations,  Moscow, Nauka, 1991 (in Russian).

\bibitem{ca2}
Casati P., Falqui G., Magri F., Pedroni, M., The KP theory
revisited I, II, III, IV, SISSA Preprint, 1996, SISSA/2-5/96/FM.

\bibitem{cr}
Cresswell C., Joshi N., The discrete first, second and
thirty-fourth Painlev\'e hierarchies, {\it J. Phys. A: Math.
Gen.}, 1999, V.32, 655--669.

\bibitem{flaschka}
Flaschka H., Newell A.C., Monodromy- and spectrum-preserving
deformations I, {\it Comm. Math. Phys.}, 1980, V.76, 65--116.

\bibitem{fokas}
Fokas A.S., Its A.R., Kitaev A.R., The isomonodromy approach to
matrix models in 2D quantum gravity, {\it Comm. Math. Phys.},
1992, V.147, 395--430.

\bibitem{fokas1}
Fokas A.S., Grammaticos B., Ramani A., From continuous to discrete
Painlev\'e equations, {\it J. Math. Anal. Appl.}, 1993, V.180,
342--360.

\bibitem{gr1}
Grammaticos B., Ramani A., Papageorgiou V., Do integrable mapping
have the Painlev\'e property?, {\it Phys. Rev. Lett.}, 1991, V.67,
1825--1828.

\bibitem{gram1}
Grammaticos B., Papageorgiou V., Ramani A., Discrete dressing
transformations and Painlev\'e equations, {\it Phys. Lett.~A},
1997, V.235, 475--479.

\bibitem{gram}
Grammaticos B., Ramani A., Satsuma J., Willox R., Carstea A.S.,
Reductions of integrable lattices, {\it J.~Nonlinear Math. Phys.},
2005, V.12, suppl.~1, 363--371.

\bibitem{kazakova}
Kazakova T.G., Finite-dimensional reductions of the discrete Toda
chain, {\it J.~Phys.~A: Math. Gen.}, 2004, V.37, 8089--8102.

\bibitem{mpz}
Magri F., Pedroni M., Zubelli J.P., On the geometry of Darboux
transformations for the KP hierarchy and its connection with the
discrete KP hierarchy, {\it Comm. Math. Phys.}, 1997, V.188,
305--325.

\bibitem{nijhoff}
Nijhoff F.W., Satsuma J., Kajiwara K., Grammaticos B., Ramani A.,
A study of the alternative discrete Painlev\'e-II equation, {\it
Inverse Problems}, 1996, V.12, 697--716.

\bibitem{nijhoff1}
Nijhoff F.W., Papageorgiou V.G., Similarity reductions of
integrable lattices and discrete analogues of the Painlev\'e II
equation, {\it Phys. Lett.~A}, 1991, V.153, 337-344.

\bibitem{noumi1}
Noumi M., Yamada Y., Affine Weyl groups, discrete dynamical
systems and Painlev\'e equations, {\it Comm. Math. Phys.}, 1998,
V.199, 281--295.

\bibitem{noumi2}
Noumi M., Yamada Y., Higher order Painlev\'e equations of type
$A_l^{(1)}$, {\em Funkcial. Ekvac.}, 1998, V.41, 483--503.

\bibitem{sv2}
Svinin A.K., Extension of the discrete KP hierarchy, {\it J. Phys.
A: Math. Gen.}, 2002, V.35, 2045--2056.

\bibitem{sv1}
Svinin A.K., Extended discrete KP hierarchy and its reductions
from a geometric viewpoint, {\it Lett. Math. Phys.}, 2002, V.61,
231--239.

\bibitem{sv5}
Svinin A.K., Invariant submanifolds of the
Darboux--Kadomtsev--Petviashvili chain and an extension of the
discrete Kadomtsev--Petviashili hierarchy, {\it Theor. Math.
Phys.}, 2004, V.141, 1542--1561.

\bibitem{ueno}
Ueno K., Takasaki K., Toda lattice hierarchy. I, II, {\it Proc.
Japan Acad. Ser. A Math. Sci.}, 1983, V.59, 167--170; 215--218.

\bibitem{veselov}
Veselov A.P., Shabat A.B., Dressing chains and the spectral theory
of the Schr\"odinger operator, {\it Funct. Anal. Appl.}, 1993,
V.27, 81--96.


\end{thebibliography}
\end{document}